\documentclass[natbib,epj]{svjour}
\usepackage{amsmath}
\usepackage{amssymb}
\usepackage{amsfonts}
\usepackage{graphicx}% Include figure files
\usepackage{dcolumn}% Align table columns on decimal point
\usepackage{bm}% bold math
\usepackage{color}% color
\usepackage{multirow}

\begin{document}
\title{Statistical properties of world investment networks}%

\author{Dong-Ming Song\inst{1,2}, Zhi-Qiang Jiang\inst{1,2}, \and Wei-Xing Zhou\inst{1,2,3,4,}
\thanks{e-mail: wxzhou@ecust.edu.cn}%
}                     % Do not remove
%
%\offprints{Wei-Xing Zhou}          % Insert a name or remove this line
%
\institute{School of Business, East China University of Science and
Technology, Shanghai 200237, China \and School of Science, East
China University of Science and Technology, Shanghai 200237, China
\and Research Center for Econophysics, East China University of
Science and Technology, Shanghai 200237, China \and Research Center
of Systems Engineering, East China University of Science and
Technology, Shanghai 200237, China}
\date{Received: \today / Revised version: }
\abstract{We have performed a detailed investigation on the world
investment networks constructed from the Coordinated Portfolio
Investment Survey (CPIS) data of the International Monetary Fund,
ranging from 2001 to 2006. The distributions of degrees and node
strengthes are scale-free. The weight distributions can be well
modeled by the Weibull distribution. The maximum flow spanning trees
of the world investment networks possess two universal allometric
scaling relations, independent of time and the investment type. The
topological scaling exponent is $1.17\pm0.02$ and the flow scaling
exponent is $1.03\pm0.01$.
 % end of PACS codes
} %end of abstract
%
%\authorrunning{D.-M. Dong, Z.-Q. Jiang, \& W.-X. Zhou}
%\titlerunning{World Investment Networks}
\maketitle

\section{Introduction}

The constituents of a complex system and their interactions can be
characterized by a complex network. The network perspective has
stimulated explosive interests in the research of social,
informational, technological, and biological systems, resulting in
deeper understanding of complex systems
\cite{Albert-Barabasi-2002-RMP,Newman-2003-SIAMR,Dorogovtsev-Mendes-2003,Boccaletti-Latora-Moreno-Chavez-Hwang-2006-PR}.
As a part of social systems, the network properties of many economic
and financial systems have been studied. This literature grows fast
and we try to present a very brief review below.

The stocks in a stock market belong to different industrial sectors.
Generally speaking, the prices of stocks in the same sector evolve
in a correlated manner. If we treat each stock as a node and the
distance of two stocks based on the cross-correlation coefficient as
the weight linking the two nodes, then the market forms a network.
The minimal spanning tree extracted from the distance matrix can be
used to investigate the hierarchical structure of a portfolio of
stocks, which is usually related to industrial sectors
\cite{Mantegna-1999-EPJB,Bonanno-Vandewalle-Mantegna-2000-PRE,Bonanno-Lillo-Mantegna-2001-QF,Onnela-Chakraborti-Kaski-Kertesz-2002-EPJB}.
Other topological properties of stock market networks are also
studied for different markets
\cite{Kim-Kim-Lee-Kahng-2002-JKPS,Tumminello-Aste-DiMatteo-Mantegna-2005-PNAS,Eom-Oh-Kim-2007-JKPS}.

Alternatively, rather than considering a portfolio stocks, the price
time series of a single stock can also be mapped into networks,
which enables us to investigate the dynamics of a stock through its
network structure. There are several methods for this purpose. For a
pseudoperiodic time series, one can partition it into disjoint
cycles according to the local minima or maxima, and each cycle is
considered a basic node of a network, in which two nodes are deemed
connected if the phase space distance or the correlation coefficient
between the corresponding cycles is less than a predetermined
threshold \cite{Zhang-Small-2006-PRL}. We note that a weighted
network can also be constructed if the phase space distance or the
correlation coefficient is treated as the weigh of a link. This
method for pseudoperiodic time series can also be generalized to
other time series, which has been applied to stock time series
\cite{Yang-Yang-2008-PA}. Other methods for network construction
from time series are based on fluctuation patterns
\cite{Li-Wang-2006-CSB,Li-Wang-2007-PA}, visibility of nodes
\cite{Lacasa-Luque-Ballesteros-Luque-Nuno-2008-PNAS}, and so on.

\begin{table*}[!htp]
 \caption{\label{Tb:NodesEdges} The number of nodes $N_v$, the number of arcs $N_a$ (directed),
 and the number of edges $N_e$ (undirected) of the world investment networks constructed from different
  kinds of investment data for different years.}
 \medskip
 \centering
 \begin{tabular}{ccccccccccccccccccccccccccccc}
 \hline \hline
  \multirow{3}*[1.5mm]{Year}&& \multicolumn{3}{c}{TP}&&\multicolumn{3}{c}{ES}&&\multicolumn{3}{c}{TD}&&\multicolumn{3}{c}{SD}&&\multicolumn{3}{c}{LD}\\  %
  \cline{3-5}  \cline{7-9} \cline{11-13} \cline{15-17} \cline{19-21}
                   && $N_v$ & $N_a$ & $N_e$ && $N_v$ & $N_a$ &$N_e$ && $N_v$ & $N_a$ &$N_e$ && $N_v$ & $N_a$ &$N_e$ && $N_v$ & $N_a$ &$N_e$ \\
    \hline
    2001           && 195   & 3244  & 2397  && 167   & 2261  & 1701 && 168   & 2655  & 1959 && 120   & 971   & 778  && 168   & 2649  & 1961 \\
    2002           && 194   & 3325  & 2477  && 172   & 2315  & 1737 && 174   & 2710  & 2042 && 136   & 1043  & 864  && 172   & 2716  & 2044 \\
    2003           && 197   & 3597  & 2649  && 177   & 2930  & 2216 && 176   & 2910  & 2193 && 134   & 1108  & 888  && 170   & 2893  & 2167 \\
    2004           && 193   & 3791  & 2774  && 174   & 3111  & 2344 && 173   & 3136  & 2339 && 133   & 1156  & 930  && 165   & 2968  & 2258 \\
    2005           && 195   & 4001  & 2926  && 173   & 2764  & 2055 && 170   & 3349  & 2482 && 130   & 1260  & 1016 && 164   & 3117  & 2367 \\
    2006           && 210   & 4483  & 3248  && 193   & 3187  & 2353 && 182   & 3592  & 2638 && 145   & 1353  & 1080 && 172   & 3415  & 2528 \\
    \hline \hline
 \end{tabular}
\end{table*}

There are also intense interests in the study of world trade webs,
which describe the international trade relations between different
economies. Serrano and Bogu{\~n}{\'a} have constructed a world trade
web utilizing the COMTRADE database of the United Nations Statistics
Division and pointed out that the world trade web exhibits typical
features of complex networks \cite{Serrano-Boguna-2003-PRE}. A
fitness model \cite{Garlaschelli-Loffredo-2004-PRL} has been
proposed to reproduce the topology of world trade webs based on a
different and more detailed data set \cite{Gleditsch-2002-JCR}, in
which the fitness of a node is associated with the GDP of the
corresponding economy. The fitness model was then extended to an
evolving and directed description of world trade webs
\cite{Garlaschelli-Loffredo-2005-PA}. Furthermore, the interrelation
between the topology of a world trade web and the GDP of countries
has been elaborated
\cite{Garlaschelli-DiMatteo-Aste-Caldarelli-Loffredo-2007-EPJB}. The
world trade web can also be described by weighted networks, in which
not only the topology but also the trade volume are considered
\cite{Fagiolo-Reyes-Schiavo-2008-PA,Bhattacharya-Mukherjee-Saramaki-Kaski-Manna-2008-JSM}.
A gravity model was used to model the weighted world trade web
\cite{Hoff-Ward-2005-XXX,Bhattacharya-Mukherjee-Saramaki-Kaski-Manna-2008-JSM}.
The world trade webs also show universal allometric scaling
\cite{Duan-2007-EPJB}, synchronization \cite{Li-Jin-Chen-2003-PA},
community structure \cite{Tzekina-Danthi-Rockmore-2008-EPJB}, and
rich-club structure
\cite{Bhattacharya-Mukherjee-Saramaki-Kaski-Manna-2008-JSM}.

There are also other economic networks, such as the world exchange
arrangements web \cite{Li-Jin-Chen-2004-PA}, the ``product space''
networks \cite{Hidalgo-Klinger-Barabasi-Hausmann-2007-Science}, the
venture capital networks
\cite{Hochberg-Ljungqvist-Lu-2007-JF,Kogut-Urso-Walker-2007-MS}, the
stock investment networks
\cite{Battiston-Rodrigues-Zeytinoglu-2007-ACS}, and the bank
connection networks
\cite{Iori-Masi-Precup-Gabbi-Caldarelli-2008-JEDC}. In this work, we
shall study the statistical properties of a new economic network
named the world investment network (WIN). The remainder of the work
is organized as follow. In Section \ref{S1:Data}, we briefly
describe the data sets adopted and construct the world investment
networks. We investigate the basic statistical properties of
unweighted networks in Section \ref{S1:uwWIN} and weighted networks
in Section \ref{S1:wWIN}. We further study the allometric scaling in
Section \ref{S1:AScaling}. Section \ref{S1:Conclusion} concludes.

\section{Constructing world investment networks}
\label{S1:Data}

We construct the world investment networks using the Coordinated
Portfolio Investment Survey (CPIS) data released by the
International Monetary Fund (IMF). The CPIS data are publicly
available and can be retrieved from the IMF web site. There are five
kinds of data including total portfolio investment (TP), equity
securities (ES), total debt securities (TD), long-term debt
securities (LD), and short-term debt securities (SD). All these data
are recorded from 2001 to 2006. Therefore, there are 30 tables in
total. Each table contains the investment information among
different economies.

\begin{figure}[htb]
\centering
\includegraphics[width=8cm]{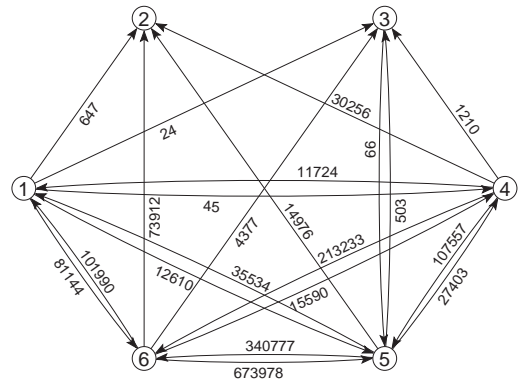}
\caption{\label{Fig:Example} Diagram illustrating part of the world
investment network obtained from ES in 2001. The numbers 1, 2, 3, 4,
5, and 6 in the open circles signify Australia, China, Egypt,
Luxembourg, United Kingdom, and United States, respectively. The
numbers above the edges represent the network weights (investment
volumes).}
\end{figure}

A directed and weighted network can be constructed from each table.
In each network, the nodes represent economies. If economy $i$
invests in economy $j$, we can draw a directed link $i\to j$ from
node $i$ to node $j$, to which a weight identical to the investment
volume is assigned. The directed and weighted network can be fully
expressed by a matrix $W$, where the element $w_{ij}$ stands for the
investment volume from economy $i$ to economy $j$. We note that
$w_{ii}=0$ for all economies and $w_{ij}=0$ if economy $i$ does not
invest in $j$. The matrix $W$ does not need to be symmetric, that
is, $w_{ij}\neq w_{ji}$. A schematic diagram is illustrated in
Figure~\ref{Fig:Example}, which is a part of the network constructed
from the ES data in 2001. When $w_{ij}=0$, there is no link from $i$
to $j$ in the figure.

The directed network $W$ can be converted into an undirected network
$G$, whose weight is determined by
\begin{equation}
 g_{ij} = g_{ji} = w_{ij}+w_{ji}~.
 \label{Eq:gij}
\end{equation}
For convenience, directed and undirected links are called arcs and
edges, respectively. Let $N_v$ be the number of nodes of network
$W$. Then, the number of arcs of $W$ is
\begin{equation}
 N_a = \sum_{i=1}^{N_v} \sum_{j=1}^{N_v} \mathbf{I}(w_{ij})~,
 \label{Eq:Na}
\end{equation}
where the indicator function $\mathbf{I}(x)$ equals to 1 if $x>0$
and 0 otherwise, and the number of edges of $G$ is
\begin{equation}
 N_e = \frac{1}{2}\sum_{i=1}^{N_v} \sum_{j=1}^{N_v} \mathbf{I}(g_{ij})
     = \sum_{i=1}^{N_v} \sum_{j=i}^{N_v} \mathbf{I}(g_{ij}) ~,
 \label{Eq:Ne}
\end{equation}
It is obvious that $2N_e\geqslant N_a$. Table~\ref{Tb:NodesEdges}
reports the values of $N_v$, $N_a$ and $N_e$ for all the 30
networks. Roughly speaking, for each of the five investment types,
the number of nodes $N_a$ is almost constant before 2006 and
increases sharply in 2006, while the number of links ($N_e$ and
$N_a$) increases gradually from 2001 to 2006 for both directed and
undirected networks, which implies an increasing globalization.

\section{Basic statistical properties of unweighted world investment networks}
\label{S1:uwWIN}

\subsection{Undirected networks}

An undirected and unweighted network $A$ can be extracted from an
undirected and weighted network $G$. We note that $A$ is the
adjacency matrix of $G$. The element $a_{ij}$ of $A$ can be
determined as follows,
\begin{equation}
 a_{ij} = {\mathbf{I}}(g_{ij})~.
\label{eq:aij}
\end{equation}
Speaking differently, $a_{ij}=1$ if economy $i$ invests in $j$ or
$j$ invests in $i$ or both, and $a_{ij}=0$ otherwise.

\begin{figure}[htb]
\centering
\includegraphics[width=8cm]{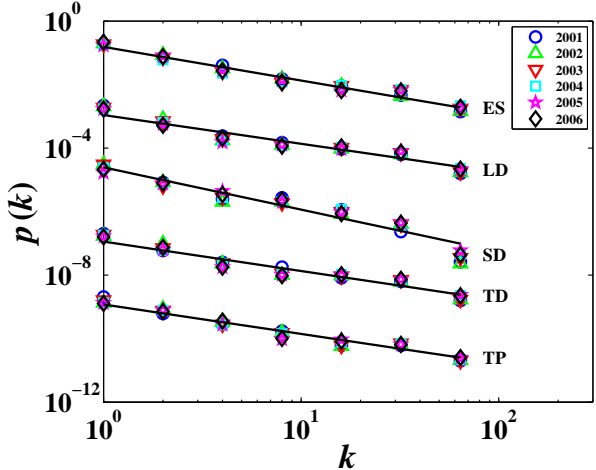}
\caption{\label{Fig:DegreeUD} (Color online) Degree distributions of
the constructed undirected WINs. The data points for LD, SD, TD, and
TP have been translated vertically by a factor of $10^{-1}$,
$10^{-2}$, $10^{-3}$, and $10^{-4}$ for clarity. The solid lines are
the best power-law fits to the data.}
\end{figure}

We first investigate the degree distribution of nodes for each
network. The degree of node $i$ can be calculated as follows
\begin{equation}
 k_i = \sum_{j=1}^{N_v} a_{ij}~.
 \label{Eq:ki}
\end{equation}
Figure~\ref{Fig:DegreeUD} shows the degree distributions for the 30
undirected networks. For clarity, the data for different types of
investment are shifted vertically. For the networks constructed from
the same investment type, the six degree distributions for different
years almost collapse onto a single curve. For the networks from
different investment types, the distributions are different from
each other.

An evident feature of the degree distributions is that they all
exhibit a power-law behavior:
\begin{equation}
p(k) \sim k^{-\gamma}.
 \label{Eq:DegreePL}
\end{equation}
It means that the world investment networks are scale-free. The
power-law exponent $\gamma$ can be estimated through a linear
least-squares regression to fit the data in log-log coordinates.
Table~\ref{Tb:gammaUN} reports the exponents $\gamma$ for all the
networks. It is found that, for each type of networks, the exponents
$\gamma$ slightly decreases along time. The observation means that
there are more highly connected economies recently, also indicating
an increase in globalization.

\begin{table}[!htp]
 \caption{\label{Tb:gammaUN} The power-law exponents $\gamma$ of the degree distributions
 for undirected and unweighted networks. The numbers in the parentheses are the errors magnified by 100.}
 \medskip
 \centering
 \begin{tabular}{cr@{}lr@{}lr@{}lr@{}lr@{}lccccc}
 \hline \hline
    Year & \multicolumn{2}{c}{TP}&  \multicolumn{2}{c}{ES} &  \multicolumn{2}{c}{TD} &  \multicolumn{2}{c}{SD} &  \multicolumn{2}{c}{LD} \\
    \hline
    2001 & 1.03&(8) & 1.12&(5)  & 1.04&(9)  & 1.46&(19) & 1.04&(10)\\
    2002 & 1.00&(8) & 1.13&(4)  & 1.04&(11) & 1.47&(20) & 1.03&(12)\\
    2003 & 1.00&(9) & 1.01&(10) & 1.02&(10) & 1.39&(16) & 0.99&(12)\\
    2004 & 0.95&(7) & 0.95&(9)  & 0.95&(10) & 1.32&(14) & 0.92&(11)\\
    2005 & 0.93&(9) & 1.04&(8)  & 0.91&(12) & 1.24&(12) & 0.90&(13)\\
    2006 & 0.93&(8) & 1.06&(10) & 0.93&(12) & 1.33&(14) & 0.91&(11)\\
    \hline \hline
 \end{tabular}
\end{table}

The average minimum path length is among the most studied quantities
of complex networks. Table~\ref{Tb:path} lists the path length of
all the networks. We see that the values of average minimum path
length are quite small. For the TP, TD and LD networks, the average
minimum path length decreases. For the ES networks, the average
minimum path length does not have a clear trend and reaches a
maximum value in 2006. For the SD networks, it is also hard to
identify an evident trend. It is noteworthy to point out that the
average minimum path length does not change much from one year to
another.

\begin{table}[htp]
 \caption{\label{Tb:path} The average path lengthes of all the undirected and unweighted networks.
  The numbers in the parentheses are the errors magnified by 100.}
 \medskip
 \centering
 \begin{tabular}{cccccc}
 \hline \hline
    Year & TP    & ES    & TD  & SD   & LD \\
    \hline
    2001 & 1.89 & 1.91 & 1.86 & 1.99 & 1.86 \\
    2002 & 1.88 & 1.94 & 1.90 & 2.12 & 1.88 \\
    2003 & 1.87 & 1.86 & 1.86 & 2.03 & 1.83 \\
    2004 & 1.84 & 1.84 & 1.84 & 2.01 & 1.81 \\
    2005 & 1.83 & 1.91 & 1.80 & 1.92 & 1.79 \\
    2006 & 1.84 & 1.95 & 1.82 & 1.97 & 1.79 \\
    \hline \hline
 \end{tabular}
\end{table}

The clustering coefficient of a node is a measure of the cluster
structure indicating how much the adjacent vertices of its adjacent
vertices are its adjacent vertices. In other words, the clustering
coefficient of node is the ratio of the number of existing edges
between its adjacent vertices to the number of possible edges
between them. Table~\ref{Tb:CluCeo} presents the average clustering
coefficients for all the networks. The average clustering
coefficients of networks with the same investment type are almost
the same for different years. A closer scrutiny shows that, the
average clustering coefficient roughly decreases along time for the
ES type and increases for the other four types.

\begin{table}[htp]
 \caption{\label{Tb:CluCeo} The average clustering coefficients of all the undirected and unweighted networks.}
 \medskip
 \centering
 \begin{tabular}{cccccc}
 \hline \hline
    Year & TP   & ES   & TD   & SD   & LD \\
    \hline
    2001 & 0.67 & 0.67 & 0.66 & 0.52 & 0.64 \\
    2002 & 0.73 & 0.66 & 0.68 & 0.49 & 0.66 \\
    2003 & 0.71 & 0.67 & 0.67 & 0.50 & 0.67 \\
    2004 & 0.73 & 0.66 & 0.66 & 0.55 & 0.66 \\
    2005 & 0.73 & 0.67 & 0.68 & 0.60 & 0.67 \\
    2006 & 0.74 & 0.64 & 0.70 & 0.60 & 0.68 \\
    \hline \hline
 \end{tabular}
\end{table}

\subsection{Directed networks}

\begin{figure}[htb]
\centering
\includegraphics[width=6.5cm]{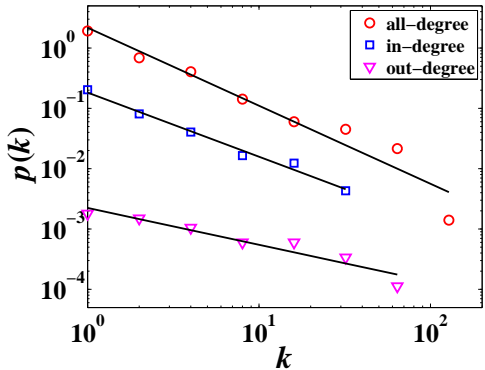}
\caption{\label{Fig:Degree} (Color online) Degree distributions of
the directed and unweighted network $B$ constructed from the ES data
in 2001. The data points for in-degree and out-degree have been
translated vertically by a factor of 0.1 and 0.01 for clarity. The
solid lines are the best power-law fits to the data.}
\end{figure}

A directed and unweighted network $B$ is the adjacency matrix of the
corresponding directed and weighted network $W$. The element
$b_{ij}$ of $B$ can be determined as follows,
\begin{equation}
 b_{ij} = {\mathbf{I}}(w_{ij})~.
\label{eq:bij}
\end{equation}
Speaking differently, $b_{ij}=1$ if economy $i$ invests in economy
$j$ and $b_{ij}=0$ otherwise. The in-degree $k^{\rm{in}}$,
out-degree $k^{\rm{out}}$, and all-degree $\gamma^{\rm{all}}$ are
defined by,
\begin{equation}
 k^{\rm{in}}_i = \sum_{j=1}^{N_v} b_{ij}~,~k^{\rm{out}}_i = \sum_{j=1}^{N_v} b_{ij}~,~k_i^{\rm{all}} =
 k^{\rm{in}}_i + k^{\rm{out}}_i~.
 \label{Eq:InOutDegree}
\end{equation}
Figure~\ref{Fig:Degree} illustrates the degree distributions of the
directed and unweigthed network $B$ constructed from the ES data in
2001.

We observe that the probability distributions of the degrees are
consistent with a power-law behavior:
\begin{equation}
 p(k^{\rm{io}}) \sim (k^{\rm{io}})^{-\gamma_{\rm{io}}},~
 \label{Eq:p:k:gammas}
\end{equation}
where ${\rm{io}}={\rm{in}}$ for in-degrees, ${\rm{io}}={\rm{out}}$
for out-degrees, and ${\rm{io}}={\rm{all}}$ for all-degrees. Linear
least-squares regressions give the estimates of the three power-law
exponents: $\gamma^{\rm{all}} = 1.17 \pm 0.07$ for all-degrees,
$\gamma^{\rm{in}} = 1.06 \pm 0.07$ for in-degrees, and $0.58 \pm
0.06$ for out-degrees, respectively. The exponents
$\gamma^{\rm{all}}$, $\gamma^{\rm{in}}$, and $\gamma^{\rm{out}}$ for
the all the directed and unweighted networks are reported in
Table~\ref{Tb:uW:D:gamma}. On average, the exponents
$\gamma^{\rm{all}}$, $\gamma^{\rm{in}}$, and $\gamma^{\rm{out}}$
decrease along time for each investment type. This shows that the
investments among economies become much denser from year to year,
also a signal of an increasing globalization from 2001 to 2006.

\begin{table}[htp]
 \caption{\label{Tb:uW:D:gamma} The power-law exponents
  $\gamma^{\rm{all}}$ (top panel), $\gamma^{\rm{in}}$ (middle panel)
  and $\gamma^{\rm{out}}$ (bottom panel) of the degree distributions
  for all the directed and unweighted networks.
  The numbers in the parentheses are the errors magnified by 100.}
 \medskip
 \centering
  \begin{tabular}{cr@{}lr@{}lr@{}lr@{}lr@{}lccccc}
 \hline \hline
    Year & \multicolumn{2}{c}{TP}&  \multicolumn{2}{c}{ES} &  \multicolumn{2}{c}{TD} &  \multicolumn{2}{c}{SD} &  \multicolumn{2}{c}{LD} \\
    \hline
    2001 & 1.05&(6) & 1.17&(7) & 1.00&(7) & 1.20&(14) & 1.00&(7) \\%
    2002 & 1.02&(5) & 1.12&(5) & 1.03&(7) & 1.19&(10) & 0.99&(8) \\%
    2003 & 1.01&(5) & 1.00&(6) & 1.00&(6) & 1.25&(12) & 0.97&(8) \\%
    2004 & 0.97&(3) & 0.98&(6) & 0.97&(7) & 1.15&(6) & 0.94&(8) \\%
    2005 & 0.97&(7) & 1.04&(6) & 0.94&(8) & 1.12&(7) & 0.94&(9) \\%
    2006 & 0.97&(6) & 1.02&(7) & 0.96&(8) & 1.22&(5) & 0.94&(7) \\%
    \hline
    2001 & 1.03&(8) & 1.06&(6) & 0.95&(8) & 1.38&(7) & 0.95&(9) \\%
    2002 & 0.83&(6) & 1.04&(7) & 1.50&(31) & 1.42&(7) & 0.85&(14) \\%
    2003 & 0.89&(7) & 1.01&(9) & 1.05&(10) & 1.37&(10) & 0.91&(11) \\%
    2004 & 0.77&(7) & 0.91&(8) & 0.88&(8) & 1.30&(7) & 0.83&(11) \\%
    2005 & 0.77&(11) & 1.08&(7) & 0.79&(13) & 1.20&(7) & 0.78&(13) \\%
    2006 & 0.73&(12) & 1.03&(8) & 0.75&(13) & 1.29&(5) & 0.71&(12) \\%
    \hline
    2001 & 0.35&(15) & 0.58&(7)  & 0.16&(15) & 0.81&(17) & 0.17&(16) \\%
    2002 & 0.42&(16) & 0.51&(10) & 0.32&(11) & 0.75&(18) & 0.31&(16) \\%
    2003 & 0.53&(24) & 0.19&(14) & 0.19&(15) & 0.85&(15) & 0.18&(14) \\%
    2004 & 0.23&(17) & 0.18&(19) & 0.17&(19) & 0.75&(11) & 0.19&(20) \\%
    2005 & 0.31&(18) & 0.15&(15) & 0.19&(16) & 0.91&(13) & 0.24&(16) \\%
    2006 & 0.05&(10) & 0.23&(8)  & 0.16&(18) & 0.64&(27) & 0.14&(27) \\%
    \hline \hline
 \end{tabular}
\end{table}

\begin{table}[htp]
 \caption{\label{Tb:uW:D:path} The average minimum path lengthes of all the directed and unweighted networks.}
 \medskip
 \centering
 \begin{tabular}{cccccc}
 \hline \hline
    Year & TP   & ES   & TD   & SD   & LD \\
    \hline
    2001 & 4.68 & 4.24 & 4.16 & 3.82 & 4.16 \\
    2002 & 4.65 & 4.48 & 4.32 & 4.58 & 4.30 \\
    2003 & 4.50 & 4.28 & 4.21 & 4.26 & 4.09 \\
    2004 & 4.28 & 4.01 & 3.94 & 4.05 & 3.85 \\
    2005 & 4.24 & 4.09 & 3.72 & 3.76 & 3.72 \\
    2006 & 4.51 & 4.46 & 4.05 & 4.13 & 3.87 \\
    \hline \hline
 \end{tabular}
\end{table}

We also report in Table~\ref{Tb:uW:D:path} the average minimum path
lengthes of the constructed directed networks. We find that, for
each type of investment, the average minimum path length gradually
decreases from 2001 to 2005, followed by a sharp increase in 2006.
This is not surprising and can be explained as follows. According to
Table \ref{Tb:NodesEdges}, the number of nodes or economies
increases abruptly in 2006, which have few links connecting to other
nodes. This considerably increases the average minimum path length
of a directed network. The increase of the average minimum path
length in 2006 does not mean a weakening globalization in 2006. On
the contrary, the inclusion of more economies in the networks
indicates a speedup in the globalization.

\section{Basic statistical properties of weighted world investment networks}
\label{S1:wWIN}

\subsection{Distribution of arc weights}

\begin{figure}[htb]
\centering
\includegraphics[width=8cm]{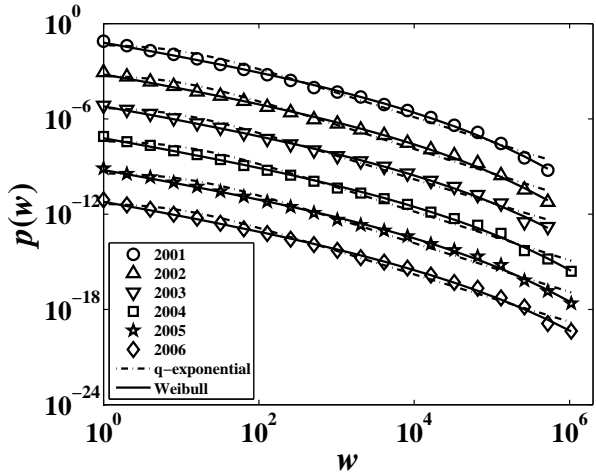}
\caption{\label{Fig:WeightPDF} Empirical probability density
functions of weights of the directed and weighted networks
constructed using the TP data from 2001 to 2006. The curves have
been shifted for clarity. The solid and dot-dashed lines are the
best fits with the Weibull and $q$-exponential distributions. }
\end{figure}

Figure~\ref{Fig:WeightPDF} plots the empirical probability
distributions of weights for the six directed and weighted networks
constructed using the TP data from 2001 to 2006. No clear evidence
of power laws is observed. We find that the results are very similar
for the ES, TD, SD and LD data (see Figure A1 in the Appendix).

We apply the Weibull and the $q$-exponential distributions to model
the weight distributions
\cite{Poloti-Scalas-2008-PA,Jiang-Chen-Zhou-2008-PA}. The Weibull
probability density $p_w(w)$ can be written as
\begin{equation}
p_w(w)  = \alpha \beta w^{\beta-1} \exp(-\alpha w^{\beta}),
 \label{Eq:WeibellDensity}
\end{equation}
and its complementary (cumulative) distribution function $C_w(w)$ is
\begin{equation}
C_w(w) = \exp(-\alpha w^{\beta}).
 \label{Eq:WeibellSurvival}
\end{equation}
When $\beta=1$, $p_w(w)$ recovers the exponential distribution. When
$0<\beta<1$, $p_w(w)$ is a stretched exponential or sub-exponential.
When $\beta>1$, $p_w(w)$ is a super-exponential. The $q$-exponential
probability density $p_q(w)$ is defined by
\begin{equation}
p_q(w)  = \mu \left[1+(1-q)(-\mu w)\right]^{\frac{q}{1-q}},
 \label{Eq:QDensity}
\end{equation}
and its complementary cumulative distribution function $C_q(w)$ is
\begin{equation}
C_q(w) = \left[1+(1-q)(-\mu w)\right]^{\frac{1}{1-q}}.
 \label{Eq:QSurvival}
\end{equation}
Usually, we have $q>1$. When $(1-q)(-\mu w)\gg1$, we observe a
power-law behavior in the tail $C_q(w)\sim w^{-{1}/{(q-1)}}$ with a
tail exponent of $1/(q-1)$.

We adopt the nonlinear least-squares estimator (NLSE) to calibrate
the weight distributions. The objective function in the fitting is
$\sum[\ln p(w)-\ln\hat p(w)]^2$ rather than $\sum[p(w)-\hat
p(w)]^2$, where $\hat p(w)$ is the empirical data. The resultant
fits are also illustrated in Figure~\ref{Fig:WeightPDF}. The
parameters of the two models and the corresponding root-mean-square
values $\chi_w$ and $\chi_q$ of the fitting residuals are shown in
Table~\ref{Tb:PDF:wij}. It is evident from
Figure~\ref{Fig:WeightPDF} that the Weibull is a better model than
the $q$-exponential, which is confirmed by the much smaller values
of $\chi_q$ compared with $\chi_w$ in Table~~\ref{Tb:PDF:wij}. We
also find that the values of $\alpha$ and $\beta$ remain constant
from 2001 to 2006. Very similar results are observed for the ES, TD,
SD and LD networks (see Table A1 and Table A2 in the Appendix).

\begin{table}[htp]
 \caption{\label{Tb:PDF:wij} Estimated parameters ($\alpha$, $\beta$, $\mu$ and
           $q$) and the RMS of fitting residuals ($\chi_w$ and $\chi_q$).}
 \medskip
 \centering
 \begin{tabular}{lllllllllllllllllllll}
 \hline \hline
  \multirow{3}*[3.2mm]{Year}&& \multicolumn{3}{c}{Weibull}&&\multicolumn{3}{c}{$q$-exponential} \\  %
  \cline{3-5}  \cline{7-9}
                   && $\alpha$ & $\beta$ & $\chi_w$ && $\mu$ & $q$ & $\chi_q$  \\
    \hline
    2001           && $0.44$ & $0.22$ & $0.0010$ && $0.05$ & $2.87$ & $0.0024$ \\
    2002           && $0.42$ & $0.23$ & $0.0006$ && $0.06$ & $2.89$ & $0.0014$ \\
    2003           && $0.43$ & $0.22$ & $0.0001$ && $0.06$ & $3.07$ & $0.0007$ \\
    2004           && $0.42$ & $0.22$ & $0.0003$ && $0.04$ & $2.87$ & $0.0018$ \\
    2005           && $0.40$ & $0.22$ & $0.0003$ && $0.04$ & $2.86$ & $0.0018$ \\
    2006           && $0.43$ & $0.21$ & $0.0004$ && $0.05$ & $3.03$ & $0.0014$ \\
    \hline \hline
 \end{tabular}
\end{table}

\subsection{Distribution of node strength}

\begin{figure}[!htb]
\centering
\includegraphics[width=6.5cm]{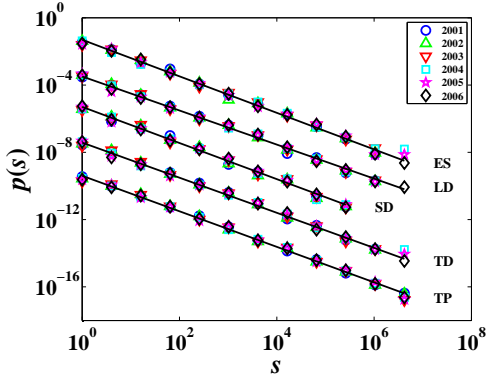}
\caption{\label{Fig:StrengthPDF}(Color online) Node strength
distributions of the directed and weighted networks. The curves have
been shifted vertically for clarity.}
\end{figure}

For weighted networks, the node strength is a generalization of the
degree, which is defined by
\begin{equation}
 s_i = \sum_{j=1}^{N_v} w_{ij} + w_{ji}~.
 \label{Eq:InOutStrength}
\end{equation}
The node strength distributions of all the directed and weighted
networks are illustrated in Figure~\ref{Fig:StrengthPDF}.

For all the node strength distributions shown in Figure
\ref{Fig:StrengthPDF}), we see nice power-law behaviors
\begin{equation}
 p(s) \sim s^{-\gamma_s}
 \label{Eq:p:s:gamma}
\end{equation}
The solid lines are the best fits to the data. The power-law
exponents $\gamma_s$ are depicted in
Table~\ref{Tb:NodeStrength:gamma}. These power-law exponents are
close to each other for each investment type. For the TP networks,
the exponents exhibit a clear decreasing trend from 2001 to 2006.

\begin{table}[htp]
 \caption{\label{Tb:NodeStrength:gamma} The power-law exponents $\gamma_s$ of the node strength distributions.
  The numbers in the parentheses are the errors magnified by 100.}
 \medskip
 \centering
 \begin{tabular}{cccccc}
 \hline \hline
    Year & TP              & ES              & TD              & SD              & LD \\
    \hline
    2001 & 1.13(4) & 1.12(3) & 1.07(2) & 1.11(4) & 1.08(3) \\%
    2002 & 1.13(4) & 1.13(7) & 1.08(2) & 1.13(4) & 1.08(2) \\%
    2003 & 1.11(4) & 1.06(2) & 1.06(2) & 1.11(4) & 1.06(2) \\%
    2004 & 1.09(3) & 1.11(4) & 1.11(4) & 1.10(3) & 1.04(2) \\%
    2005 & 1.08(3) & 1.13(4) & 1.10(4) & 1.07(3) & 1.02(2) \\%
    2006 & 1.07(2) & 1.13(4) & 1.10(4) & 1.08(2) & 1.09(4) \\%
    \hline \hline
 \end{tabular}
\end{table}

\section{Universal allometric scaling laws}
\label{S1:AScaling}

\subsection{Unweighted world investment networks}

Allometric scaling laws are ubiquitous in networking systems such as
metabolism of organisms and ecosystems river networks, food webs,
and so on
\cite{West-Brown-Enquist-1997-Science,Enquist-Brown-West-1998-Nature,West-Brown-Enquist-1999-Science,Enquist-West-Charnov-Brown-1999-Nature,Banavar-Maritan-Rinaldo-1999-Nature,Enquist-Economo-Huxman-Allen-Ignace-Gillooly-2003-Nature,Garlaschelli-Caldarelli-Pietronero-2003-Nature}.
For economic systems, the world trade webs also exhibits a universal
allometric scaling \cite{Duan-2007-EPJB}. For a complex network, a
minimal spanning tree can be extracted. Each node of the tree is
assigned a number 1. Two values $A_i$ and $C_i$ are defined for each
node $i$ in a recursive manner as follows:
\begin{equation}
 A_i = \sum_{j\in{\mathbf{J}}(i)} A_j + 1~,
 \label{Eq:A}
\end{equation}
and
\begin{equation}
 C_i = \sum_{j\in{\mathbf{J}}(i)} C_j +A_i~,
 \label{Eq:A}
\end{equation}
where ${\mathbf{J}}(i)$ stands for the set of daughter nodes of $i$
and $A_1=C_1=1$ \cite{Banavar-Maritan-Rinaldo-1999-Nature}. The
allometric scaling relation is then highlighted by the power law
relation between $C_i$ and $A_i$:
\begin{equation}
 C \sim A^{\eta}~.
 \label{Eq:ACeta}
\end{equation}
We note that not all trees have such allometric scaling, such as the
classic Cayley trees \cite{Jiang-Zhou-Xu-Yuan-2007-AICHEJ}.

For spanning trees extracted from transportation networks, the power
law exponent $\eta$ is a measure of transportation efficiency
\cite{Banavar-Maritan-Rinaldo-1999-Nature,Garlaschelli-Caldarelli-Pietronero-2003-Nature}.
The smaller is the value of $\eta$, the more efficient is the
transportation. Any spanning tree can range in principle between two
extremes, that is, the chain-like trees and the star-like trees. A
chain tree has one root and one leaf with no branching. For
chain-like trees, it is easy to show that $A_i=i$ and
$C_i=i(i+1)/2$. Asymptotically, we have a power between $C_i$ and
$A_i$ with the exponent $\eta=2^-$. For star-like trees of size $n$,
there are one root and $n-1$ leaves directly connected to the root.
We have $A=C=1$ for all the leaves and $A=n$ and $C=2n-1$ for the
root. It follows approximately that $\eta=1^+$. Therefore, if the
relation (\ref{Eq:ACeta}) holds, the exponent should be $1< \eta <
2$.

\begin{figure}[htb]
\centering
\includegraphics[width=6.5cm]{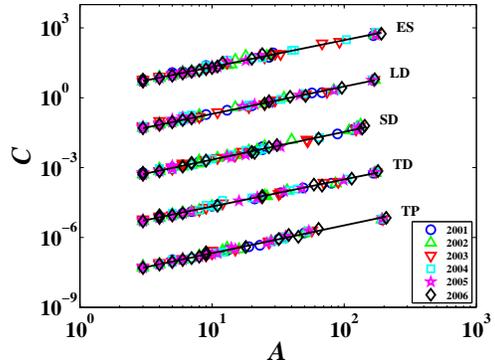}
\caption{\label{Fig:Allometric} (Color online) Allometric scaling
relationship between $S$ and $A$. The data points for LD, SD, TD,
and TP are translated vertically for clarity. The solid lines are
the best power-law fits to the data.}
\end{figure}

For each undirected and unweighted network, a minimal spanning tree
can be obtained. The calculated $C$ is plotted in
Fig.~\ref{Fig:Allometric} as a function of $A$ for each network.
Nice power-law relations are observed between $C$ and $A$. The
points $(A=1,C=1)$ for the leaves are not shown
\cite{Garlaschelli-Caldarelli-Pietronero-2003-Nature}. For each
investment type, the data points of the six networks collapse onto a
single curve.

Nice power-law behaviors are observed in Figure
\ref{Fig:Allometric}. A linear fit of $\ln C$ against $\ln A$ gives
the estimate of $\eta$ for each network. The trivial point
$(A=1,C=1)$ should be excluded from the fitting
\cite{Garlaschelli-Caldarelli-Pietronero-2003-Nature}. The resulting
exponents are listed in Table~\ref{Tb:ACeta}. We find that the
exponents are almost the same for different investment types and
different years, which means that the power-law allometric scaling
is universal for the world investment networks. This value of $\eta$
is comparable to $\eta=1.13\sim1.16$ for food webs
\cite{Garlaschelli-Caldarelli-Pietronero-2003-Nature}, but much
smaller than $\eta = 1.3$ for world trade webs \cite{Duan-2007-EPJB}
and $\eta=1.5$ for river networks
\cite{Banavar-Maritan-Rinaldo-1999-Nature}.

\begin{table}[htp]
 \caption{\label{Tb:ACeta} The topological scaling exponents $\eta$  for all the undirected and unweighted networks.
  The numbers in the parentheses are the errors magnified by 100.}
 \medskip
 \centering
 \begin{tabular}{cccccc}
 \hline \hline
    Year & TP              & ES              & TD              & SD              & LD \\
    \hline
    2001 & 1.15(3) & 1.10(4) & 1.16(1) & 1.19(2) & 1.16(2) \\%
    2002 & 1.14(3) & 1.15(4) & 1.16(2) & 1.27(3) & 1.15(2) \\%
    2003 & 1.18(3) & 1.16(2) & 1.16(2) & 1.20(3) & 1.13(3) \\%
    2004 & 1.18(3) & 1.17(3) & 1.17(3) & 1.21(2) & 1.14(3) \\%
    2005 & 1.18(3) & 1.14(3) & 1.17(2) & 1.18(3) & 1.13(3) \\%
    2006 & 1.19(3) & 1.15(3) & 1.16(2) & 1.19(2) & 1.18(2) \\%
    \hline
    mean & 1.17(2) & 1.15(2) & 1.16(1) & 1.21(3) & 1.15(2) \\%
    \hline \hline
 \end{tabular}
\end{table}

\subsection{Weighted world investment networks}

When studying the world trade webs, Duan proposed a framework of
flow allometric scaling analysis for weighted networks
\cite{Duan-2007-EPJB},
\begin{equation}
fA_i = \sum_{j\in{\mathbf{J}}(i)} fA_j + w_i~,\label{Eq:fA}
\end{equation}
and
\begin{equation}
 fC_i = \sum_{j\in{\mathbf{J}}(i)} fC_j +
fA_i,\label{Eq:fC}
\end{equation}
where ${\mathbf{J}}(i)$ is the set of daughter nodes of $i$, $w_i$
stands for the total weight flowed into node $i$, $fA_i$ is the
weighted quantity of resources and $fC_i$ stands for the weighted
transferring cost. For the leaves, the carried values of $fA$ and
$fC$ are identical to the investment volumes. One can also expect
for certain network that the allometric scaling relation exists
between $fC_i$ and $fA_i$:
\begin{equation}
 fC \sim fA^{\zeta}~,
 \label{Eq:fAfCzeta}
\end{equation}
in which the exponent $\zeta$ is called the flow allometric scaling
exponent \cite{Duan-2007-EPJB}.

\begin{figure}[htb]
\centering
\includegraphics[width=6.5cm]{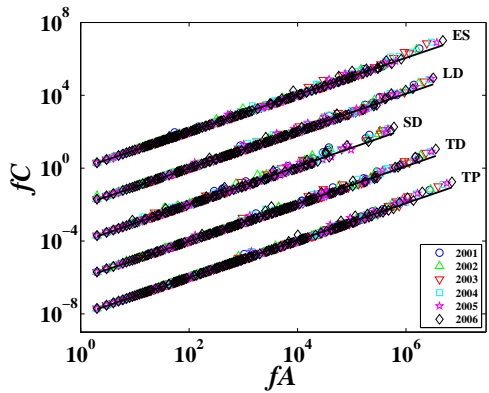}
\caption{\label{Fig:FAllometric} (Color online) Allometric scaling
relationship between $fS$ and $fA$. The data points for LD, SD, TD,
and TP are translated vertically for clarity. The solid lines are
the best power-law fits to the data.}
\end{figure}

We adopt this analysis on the maximum-flow spanning trees of the
undirected and weighted world investment networks. The maximum-flow
spanning trees of the investment networks can be obtained
\cite{Duan-2007-EPJB}. Figure~\ref{Fig:FAllometric} plots  $fC$ with
respect to $fA$ in double logarithmic coordinates. For each
investment type, the data points of the six networks collapse onto a
single curve, independent of the time.

Evident power-law behaviors are observed between $fC$ and $fA$ for
all the weighted networks under investigation. The curves seem
parallel for different types of networks. The flow scaling exponents
$\zeta$ are estimated by the slopes of the linear fits of $\ln fC$
with respect to $\ln fA$. The exponents $\zeta$ are reported in
Table~\ref{Tb:fACzeta}. We find that the flow allometric scaling
exponents are almost the same for different investment types and
different years, which means that the power-law allometric scaling
is universal for the world investment networks. It is interesting to
point out that this $\zeta$ value of the world investment networks
is close to the flow allometric scaling exponent of the world trade
webs \cite{Duan-2007-EPJB}.

\begin{table}[htp]
 \caption{\label{Tb:fACzeta} The flow scaling exponents $\zeta$  for all the undirected and weighted networks.
  The numbers in the parentheses are the errors magnified by 100.}
 \medskip
 \centering
 \begin{tabular}{cccccc}
 \hline \hline
    Year & TP              & ES              & TD              & SD              & LD \\
    \hline
    2001 & 1.02(0) & 1.01(0) & 1.03(0) & 1.07(1) & 1.03(0) \\%
    2002 & 1.02(0) & 1.01(0) & 1.02(0) & 1.06(1) & 1.02(0) \\%
    2003 & 1.02(0) & 1.03(0) & 1.03(1) & 1.05(1) & 1.02(0) \\%
    2004 & 1.02(0) & 1.03(0) & 1.03(0) & 1.04(1) & 1.03(1) \\%
    2005 & 1.02(0) & 1.01(0) & 1.03(0) & 1.04(1) & 1.03(1) \\%
    2006 & 1.02(0) & 1.01(0) & 1.03(0) & 1.04(1) & 1.02(0) \\%
    \hline
    mean & 1.02(0) & 1.02(1) & 1.03(0) & 1.05(1) & 1.03(1) \\%
    \hline \hline
 \end{tabular}
\end{table}

\section{Conclusion}
\label{S1:Conclusion}

In this work, we have constructed a new type of economic networks
based on the Coordinated Portfolio Investment Survey data released
by the International Monetary Fund, which records data from 2001 to
2006. We have studied the statistical properties of these world
investment networks. Our results show that there is an increasing
globalization in the past few years under investigation.

The degree distributions are scale-free for all the constructed
networks. For the same investment type of data, the average path
length and average clustering coefficient are almost the same for
different years with a weak trend. When we regard the world
investment networks as weighted networks, the Weibull and
$q$-exponential distributions are utilized to fit the weight
distribution by means of a nonlinear least-squares estimator. We
find that the Weibull model remarkably outperforms the
$q$-exponential model. In addition, the node strength distributions
are found to exhibit nice power-law behaviors.

We also investigated the allometric scaling of the minimal spanning
trees and the maximum-flow spanning trees of the world investment
networks. There are two universal allometric scaling exponents
characterizing the topological structure and the investment pattern
of the networks. We find that the topological scaling exponent is
$\eta=1.17\pm0.02$ and the flow scaling exponent is
$\zeta=1.03\pm0.01$. The topological scaling exponent is found to be
close to that of the food webs and smaller than that of the world
trade webs, while the flow scaling exponent is comparable to that of
the world trade webs.

\bigskip
{\textbf{Acknowledgments:}}

We are grateful to Liang Guo for helpful discussions. This work was
partly supported by the National Natural Science Foundation of China
(Grant No. 70501011), the Fok Ying Tong Education Foundation (Grant
No. 101086), and the Program for New Century Excellent Talents in
University (Grant No. NCET-07-0288).

%\pagebreak
\bibliographystyle{epj}
\bibliography{E:/Papers/Auxiliary/Bibliography}

%\end{document}

\clearpage
%\newpage
%\newpage

%\section*{Appendix}
%
%Here we provide a figure and two tables.

\setcounter{figure}{0}
\renewcommand\thefigure{A\arabic{figure}}

\begin{figure*}[htb]
\centering
\includegraphics[width=8cm]{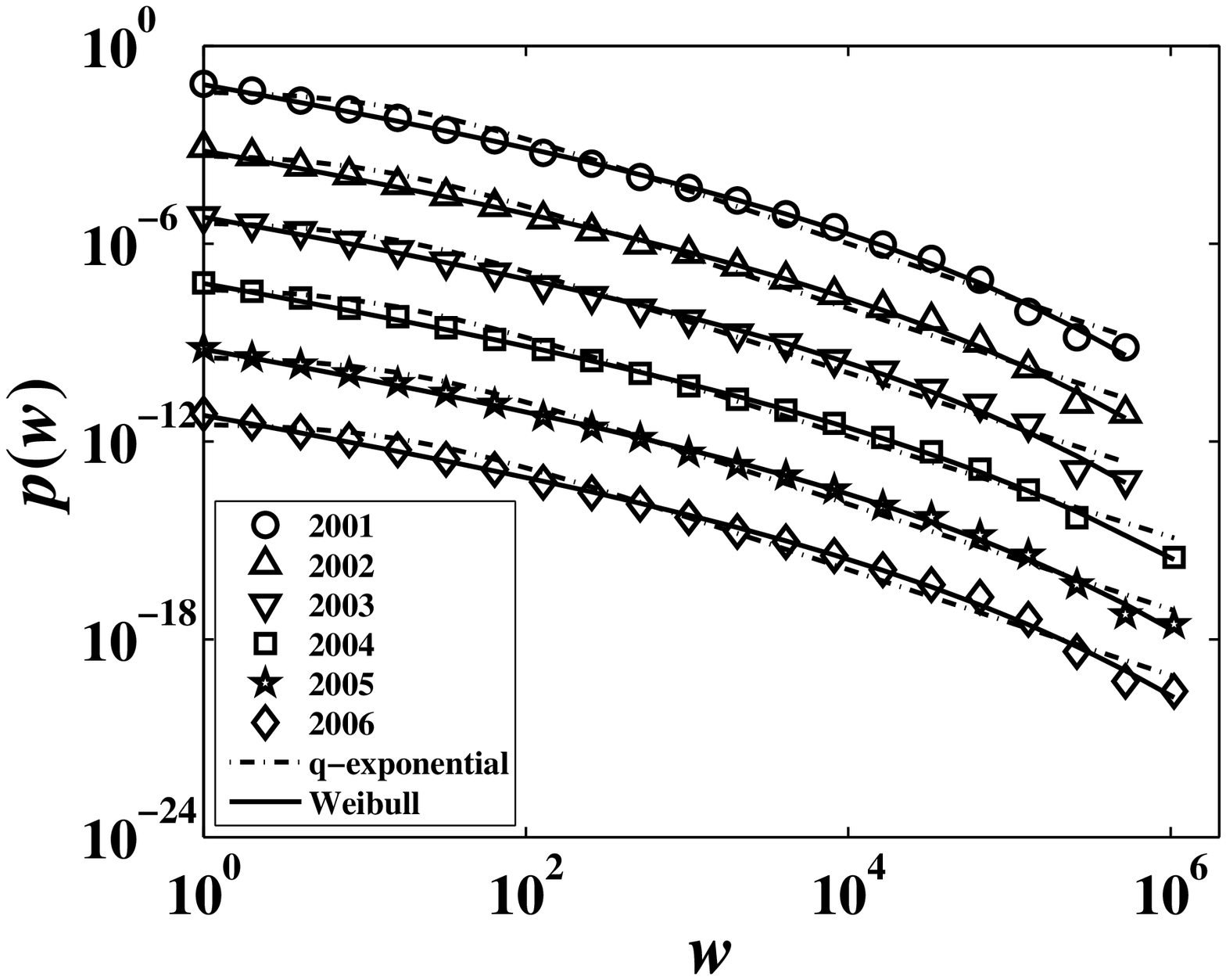}
\includegraphics[width=8cm]{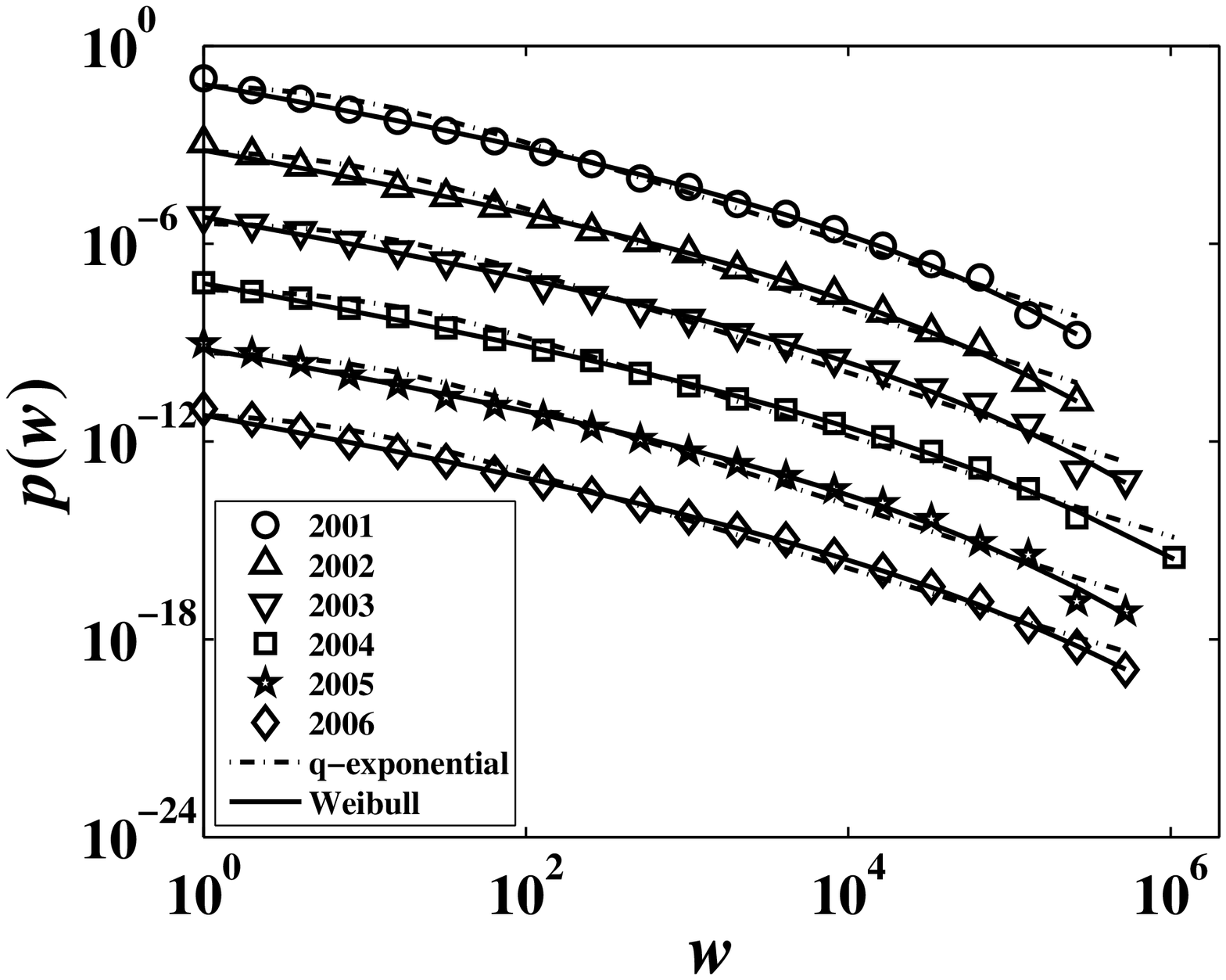}
\includegraphics[width=8cm]{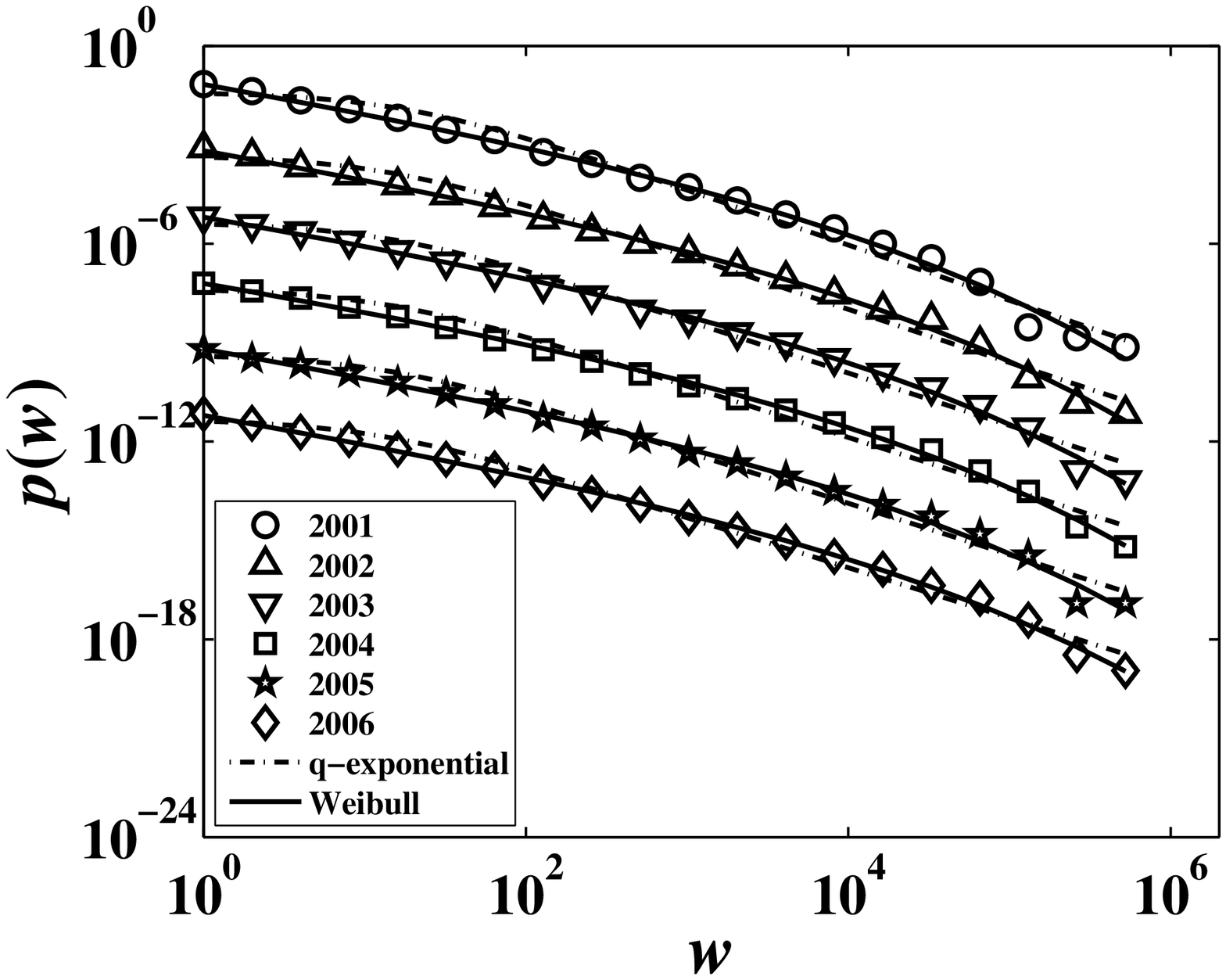}
\includegraphics[width=8cm]{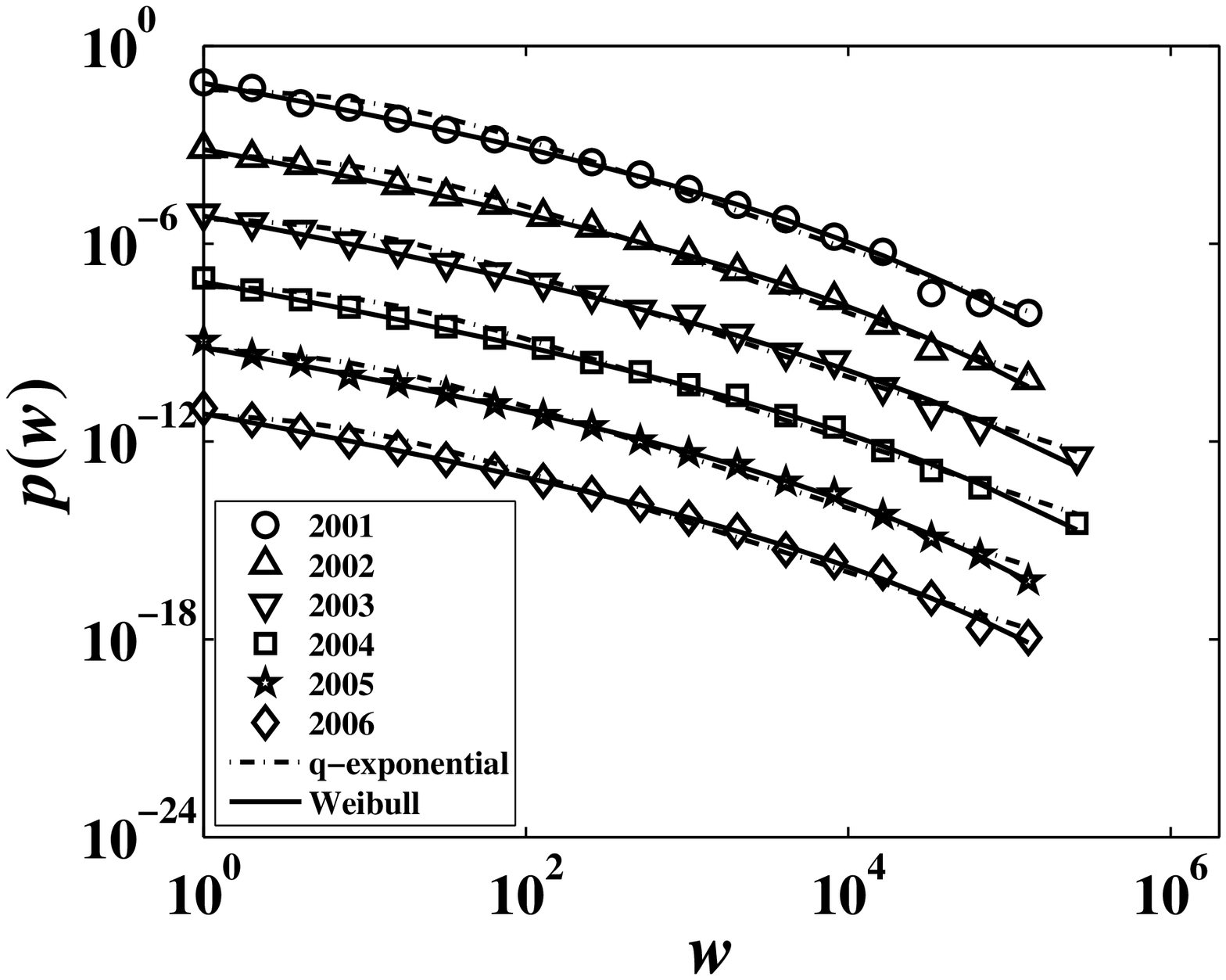}
\caption{\label{Fig:App:WeightPDF} Empirical probability density
function of weights in the constructed networks for different kinds
of investments and different years. The solid and dash-dotted lines
are the maximum likelihood fits to the Weibull and $q$-exponential
distributions, respectively.}
\end{figure*}

\setcounter{table}{0}
\renewcommand\thetable{A\arabic{table}}

\begin{table*}[!htp]
 \caption{\label{Tb:APP:wblfit} Estimated values of parameters ($\alpha$, $\beta$) by means of NLSE.
          The values of $\chi_w$ have been multiplied by 100.}
 \medskip
 \centering
 \begin{tabular}{cccccccccccccccccccccccccccc}
 \hline \hline
  \multirow{3}*[3.2mm]{Year}&& \multicolumn{3}{c}{TP}&&\multicolumn{3}{c}{ES}&&\multicolumn{3}{c}{TD}&&\multicolumn{3}{c}{SD}&&\multicolumn{3}{c}{LD}\\  %
  \cline{3-5}  \cline{7-9} \cline{11-13} \cline{15-17} \cline{19-21}
                   && $\alpha$ & $\beta$ & $\chi_w$ && $\alpha$ & $\beta$ & $\chi_w$ && $\alpha$ & $\beta$ & $\chi_w$ && $\alpha$ & $\beta$ & $\chi_w$ && $\alpha$ & $\beta$ & $\chi_w$ \\
    \hline
    2001           && $0.44$ & $0.22$ & $0.10$ && $0.46$ & $0.23$ & $0.13$ && $0.40$ & $0.23$ & $0$      && $0.38$ & $0.24$ & $0.01$ && $0.46$ & $0.22$ & $0.17$\\
    2002           && $0.42$ & $0.23$ & $0.06$ && $0.46$ & $0.22$ & $0.17$ && $0.42$ & $0.23$ & $0.04$ && $0.42$ & $0.23$ & $0.03$ && $0.42$ & $0.23$ & $0.04$\\
    2003           && $0.43$ & $0.22$ & $0.01$ && $0.38$ & $0.24$ & $0.01$ && $0.39$ & $0.23$ & $0.01$ && $0.41$ & $0.23$ & $0.02$ && $0.39$ & $0.23$ & $0.01$\\
    2004           && $0.42$ & $0.22$ & $0.03$ && $0.40$ & $0.23$ & $0$      && $0.40$ & $0.23$ & $0$      && $0.41$ & $0.23$ & $0.02$ && $0.40$ & $0.23$ & $0.01$\\
    2005           && $0.40$ & $0.22$ & $0.03$ && $0.44$ & $0.22$ & $0.10$ && $0.42$ & $0.22$ & $0$      && $0.44$ & $0.22$ & $0.04$ && $0.43$ & $0.22$ & $0$     \\
    2006           && $0.43$ & $0.21$ & $0.04$ && $0.42$ & $0.22$ & $0.12$ && $0.40$ & $0.22$ & $0.01$ && $0.44$ & $0.22$ & $0.04$ && $0.42$ & $0.22$ & $0.01$\\
    \hline \hline
 \end{tabular}
\end{table*}

\begin{table*}[!htp]
 \caption{\label{Tb:App:qfit} Estimated values of parameters ($\mu$, $q$) by means of NLSE.
          The values of $\chi_w$ have been multiplied by 100.}
 \medskip
 \centering
 \begin{tabular}{lllllllllllllllllllll}
 \hline \hline
  \multirow{3}*[3.2mm]{Year}&& \multicolumn{3}{c}{TP}&&\multicolumn{3}{c}{ES}&&\multicolumn{3}{c}{TD}&&\multicolumn{3}{c}{SD}&&\multicolumn{3}{c}{LD}\\  %
  \cline{3-5}  \cline{7-9} \cline{11-13} \cline{15-17} \cline{19-21}
                   && $\mu$ & $q$ & $\chi_q$ && $\mu$ & $q$ & $\chi_q$ && $\mu$ & $q$ & $\chi_q$ && $\mu$ & $q$ & $\chi_q$ && $\mu$ & $q$ & $\chi_q$  \\
    \hline
    2001           && $0.05$ & $2.87$ & $0.24$ && $0.07$ & $2.80$ & $0.22$ && $0.04$ & $2.74$ & $0.11$ && $0.05$ & $2.91$ & $0.07$ && $0.06$ & $2.84$ & $0.30$\\
    2002           && $0.06$ & $2.89$ & $0.14$ && $0.06$ & $2.84$ & $0.30$ && $0.06$ & $2.88$ & $0.14$ && $0.05$ & $2.77$ & $0.17$ && $0.05$ & $2.82$ & $0.15$\\
    2003           && $0.06$ & $3.07$ & $0.07$ && $0.05$ & $2.91$ & $0.07$ && $0.05$ & $2.91$ & $0.07$ && $0.05$ & $2.86$ & $0.10$ && $0.05$ & $2.87$ & $0.07$\\
    2004           && $0.04$ & $2.87$ & $0.18$ && $0.04$ & $2.74$ & $0.11$ && $0.04$ & $2.73$ & $0.10$ && $0.05$ & $3.03$ & $0.10$ && $0.05$ & $3.08$ & $0.06$\\
    2005           && $0.04$ & $2.86$ & $0.18$ && $0.05$ & $2.88$ & $0.24$ && $0.04$ & $2.84$ & $0.10$ && $0.06$ & $3.09$ & $0.13$ && $0.05$ & $3.08$ & $0.07$\\
    2006           && $0.05$ & $3.03$ & $0.14$ && $0.05$ & $2.97$ & $0.25$ && $0.04$ & $2.84$ & $0.13$ && $0.06$ & $3.17$ & $0.13$ && $0.05$ & $3.18$ & $0.08$\\
    \hline \hline
 \end{tabular}
\end{table*}

\end{document}